\documentstyle[prb,aps,preprint]{revtex}
\begin{document}
\draft
\def\zt{Zlatko Te\v sanovi\' c}
\title{\bf
Superconductivity in High Magnetic Field:
\\
Excitation Spectrum and Tunneling Properties}

\author{
 Sa\v {s}a Dukan$^{1}$
 and Zlatko Te\v {s}anovi\' {c}}
\address{
\sl Department of Physics and Astronomy,
Johns Hopkins University,
Baltimore, MD 21218, USA}
\maketitle

\begin{abstract}
The quasiparticle excitation spectrum of a type-II superconductor placed
in high  magnetic field is shown to be gapless.
 The gap turns to zero
at the points in the Magnetic Brillouin Zone which are in correspondence with
the vortex
lattice in real space.
When the field decreases below certain critical value, branch crossings occur
and gaps start opening up at the Fermi
surface.
The strong dispersion around the gapless points leads to algebraic temperature
dependence in the thermodynamic functions and the algebraic voltage dependence
in the tunneling conductance between the microscope tip and superconductor in
an STM experiment.

\end{abstract}
\pacs
{PACS numbers: 74.60.-w, 74.70.Jm}
\narrowtext

{\bf I. MOTIVATION}

The properties of a superconductor placed in an external  magnetic field have
been
a subject of considerable interest for a long time. Particularly significant in
this contest is
the mixed phase of type-II superconductor in which the external field can
coexist with superconductivity in the form of quantized flux lattice. Discovery
of high temperature superconductors (HTS), exhibiting strongly type-II
behavior, has only fueled further intense studies of
such systems.
The familiar Abrikosov-Gorkov (AG) microscopic theory based on
the semiclassical approximation \cite{gorkov} for magnetic field yields a very
good description of most conventional type-II materials. This theory completely
neglects Landau level
quantization of electronic energies in magnetic field which is justified for a
range of fields and temperatures such that $\hbar \omega _{c}\ll k_{B}T$ in a
clean system ($\hbar \omega _{c}=eH/mc$). For large impurity concentrations
this condition translates to $\omega _{c}\ll 2\pi \tau ^{-1}$, where $\tau $ is
the
scattering lifetime. Under these conditions electrons occupy a huge number of
closely separated Landau levels so that either temperature or impurity
scattering completely erases the significance of the quantized energy levels in
magnetic field.  Recently Te\v{s}anovi\'{c} et. al. examined the opposite limit
to the one described by AG theory and discovered that the inclusion of Landau
levels leads
to reentrant behavior at high fields \cite{zbt} where the superconductivity is
enhanced by magnetic field.
This behavior is mostly pronounced in the low-carrier-density systems where the
high-field limit can be readily achieved by application of fields in $1-30$
Tesla range. It is well known that HTS are {\sl inherently} strongly type-II
systems, i.e. their behavior in magnetic field is not the consequence of doping
the materials by impurities. Instead, they are actually quite clean systems
with the strong type-II behavior being due to their low carrier densities.
 There is a sharply defined Landau level structure in
such a superconductor and it should be included in any complete study of HTS in
magnetic field.

Particularly interesting in this contest is the problem of the quasiparticle
excitation spectrum in the mixed state of the superconductor.
Recent scanning-tunneling microscope (STM) experiments \cite{hess} have
revealed the
local distribution of quasiparticle states in the vortex core. Several
theoretical works have followed \cite{dorsey,gygi,klein} explaining
experimental results of Hess and
co-workers, based on the solution of the Bogoliubov-deGennes (BdG) equations
for the quasiparticle excitation spectrum in the isolated vortex case. This
situation is obtained when the external magnetic field is rather low (typically
$10^{-2}-10^{-3}$) Tesla so that vortices are well separated. It is natural to
inquire what would be the result of such an STM experiment at higher fields
($>1$ Tesla). Such an experiment would probe the electronic structure of the
vortex lattice since the STM probe would be able to scan more than just a
single vortex. A clean HTS sample would be an excellent candidate for such an
experiment: at low temperature and high fields the mean free path of the
electrons in these systems will become much longer than the separation of the
vortices. Therefore, the quasiparticle excitations will propagate {\sl
coherently}
through many unit cells of the vortex lattice. This coherent propagation will
lead to the novel features of the STM pattern.

In Section II of this paper we present the solution of BdG equations for
 the quasi-particle excitation spectrum
of the vortex lattice. We show that the type-II superconductor in high magnetic
field has a gapless excitation spectrum with the strong dispersion around
gapless points at the Fermi surface. We also discuss the mechanism of the gap
 opening in lower fields.
 In section III we discuss the behavior of the thermodynamic properties
 and the density of states for such a gapless superconductor.
 In section IV we show that the $S-N-S$ tunneling conductance and the STM
 conductance of the superconductor in high magnetic field have an algebraic
 voltage dependence.
\\\\
{\bf II. THE QUASIPARTICLE EXCITATION SPECTRUM}

We consider a 3D weakly interacting electronic system in magnetic
field with the model interaction $V(\vec {r_{1}} ,\vec {r_{2}})=-V\delta
(\vec{r_{1}}
-\vec{r_{2}})$, arising from the electron-phonon and electron-electron
pairing mechanism. We assume $V$ only weakly dependent on magnetic field.
 The Hartree-Fock Hamiltonian for such a system is:
\begin{equation}
H_{HF}=\frac{1}{2m} \sum_{\alpha} \int \Psi^{\dag}_{\alpha}(\vec{r})
(-i \hbar \nabla +\frac{e}{c}{\bf A})^{2} \Psi_{\alpha}(\vec{r})
d^{3}r + \int \Delta(\vec{r}) \Psi^{\dag}_{\uparrow}(\vec{r}) \Psi^{\dag}_
{\downarrow}(\vec{r}) d^{3}r + h.c.~~,
\label{hf}
\end{equation}
where $\Delta(\vec{r})$ is the superconducting order parameter given
by the self-consistent equation
\begin{equation}
\Delta(\vec{r})=V<\Psi_{\uparrow}(\vec{r})
\Psi_{\downarrow}(\vec{r})>
\label{ord}
\end{equation}
We take the order parameter to be uniform along the field direction.
 In the Mean Field (MF) approximation $\Delta (\vec{r})$ forms the
Abrikosov lattice of vortices that in Landau gauge $\vec{A}=H(-y,0,0)$ has
the form \cite{abr} :
\begin{equation}
\Delta _{0}(\vec{r})=\Delta \sum_{n} \exp {(\pi \frac{b_{x}}{a}n^{2})}
 \exp {[i2\pi nx/a-(y/l_{H}+
\pi nl_{H}/a)^{2}]}
\label{del}
\end{equation}
where $l_{H}=\sqrt{\hbar c/eH}$ is the magnetic length and $\Delta $ is the
 amplitude. The vortex lattice is characterized by unit vectors
 ${\vec{a}}=(a,0,0,)$ and ${\vec{b}}=(b_{x}, b_{y},0)$ ($b_x =0$, $b_{y}=a$
 for quadratic lattice and $b_x=\frac{1}{2}a$, $b_y=\frac{\sqrt{3}}{2}a$ for
triangular lattice). The flux through the unit cell is given by the expression
$ab_{y}=\pi l_{H}^{2}$.
The above form of the order parameter is entirely contained in
the lowest Landau level of the Cooper charge $2e$ and represents the excellent
 approximation as long as the region of interest is close to the $H_{c2}(T)$
 line in a phase diagram.
In lower fields, $H\approx H_{c2}(T)/(2j+1)$, $\Delta(\vec{r})$ contains
 contributions from higher Landau levels ($j\geq 1$) that
can be easily obtained by the action of the operator \cite{eil}:
\begin{equation}
\Pi ^{\dagger}(\vec{r})  = \frac{l_H}{2}(-i\frac{\partial}{\partial x}-
\frac{\partial}{\partial y} + 2\frac{y}{l_{H}^2})
\label{pi}
\end{equation}
$j$ times on $\Delta _{0}(\vec{r})$ yielding:
\begin{equation}
\Delta_{j}(\vec{r})=\Delta_{j} \sum_{n} \exp{(i\pi \frac{b_{x}}{a}n^2)}
 \exp{[i2 \pi n \frac{x}{a}-(\frac{y}{l_H}+\frac{\pi n}{a}l_H)^2]}
 H_{j}\left[ \sqrt{2} (\frac{y}{l_H}
+\frac{\pi n}{a}l_{H})\right]
\label{delj}
\end{equation}
$H_{j}(x)$ is the Hermite polynomial of the order $j$. Amplitudes $\Delta_{j}$
 can be generated from the self consistent equation (\ref{ord}). Because of the
 symmetry, only $\Delta_{j=4k}(\vec{r})$ for quadratic and
 $\Delta_{j=6k}(\vec{r})$ for triangular lattice will contribute to the
general form
of the order parameter $\Delta(\vec{r})=\sum_{j} \Delta _{j}(\vec {r})$.
 Only these functions have the same position and vorticity of zeros to produce
 the order parameter of the correct symmetry. Nevertheless, we can show that
 the higher level contributions ($k\neq 0$) are not crucial for our problem.
 They introduce extra wiggles in the spatial dependence of the order parameter
 far from the vortex positions and do not bring any new essential features in
 the form of the excitation spectrum.

In order to diagonalize the Hamiltonian (\ref{hf}), we use the
 Magnetic Sublattice Representation \cite{bychkov} with basis functions
characterized by
quasi-momentum
$\vec{q}$ perpendicular to the direction of the field. Since the electronic
 charge is half as large as the Cooper pair charge , we
choose the electronic unit cell spanned by vectors $2\vec{a}$ and
$\vec{b}$ in order to enclose one full electronic flux in the unit cell.
Then, the MBZ is defined by vectors
$\vec{a}^{*}=(b_{y}/l_{H}^{2},-b_{x}/l_{H}^{2})$ and
$\vec{b}^{*}=(0,2a/l_{H}^{2})$.
 In this basis we can write the BdG transformations \cite{deGennes} as:
\begin{eqnarray}
\Psi_{\uparrow}(\vec{r})=\sum_{k_{z},\vec{q},n}[u_{k_{z}
,\vec{q},n}c_
{\uparrow k_{z},\vec{q},n}
-v^{\dag}_
{-k_{z},-\vec{q},n}
c^{\dag}_
{\downarrow -k_{z},-\vec{q},n}
]\phi_
{k_{z},\vec{q},n}
(\vec{r}) \nonumber   \\
\Psi_{\downarrow}(\vec{r})=\sum_{k_{z},\vec{q},n}[u_{k_{z}
,\vec{q},n}c_
{\downarrow k_{z},\vec{q},n}
+v^{\dag}_
{-k_{z},-\vec{q},n}
c^{\dag}_
{\uparrow -k_{z},-\vec{q},n}
]\phi_
{k_{z},\vec{q},n}
(\vec{r})
\label{btransf}
\end{eqnarray}
where
$\phi_
{k_{z},\vec{q},n}
(\vec{r})$
are the eigenfunctions of the Magnetic Translation Group (MTG)
in the Landau gauge belonging to the
$n$-th Landau level:
\begin{eqnarray}
\phi_
{k_{z},\vec{q},n}
(\vec{r})=\frac{1}{\sqrt{2^{n}n!\sqrt{\pi}l_{H}}}
\sqrt{\frac{b_{y}}{L_{x}L_{y}L_{z}}}\exp{(ik_{z}\zeta)}
\sum_{m}\exp{(i\frac{\pi b_{x}}{2a} m^{2}-imq_{y}b_{y})}
\nonumber\\
\exp{(i(q_{x}+\frac{\pi m}{a})x
-1/2(\frac{y}{l_{H}}+
(q_{x}+\frac{\pi m}{a})l_{H
})^{2})}
H_{n}(\frac
{y}{l_{H}}+(q_{x}+\frac{\pi m}{a})l_{H}).
\label{phi}
\end{eqnarray}
$\zeta$ is the spatial coordinate
and $k_{z}$ is the momentum along the field direction.
 Product $L_{x}L_{y}L_{z}$ is the volume of the system.

The Cooper pairs are formed from the electrons having opposite crystalline
momenta $\vec{q}$ and spins within the same Landau level
(diagonal pairing) and from the electrons belonging to the Landau levels
separated by $\hbar \omega _{c}$ or more (off-diagonal
pairing). In sufficiently high magnetic fields so that $\Delta \ll \hbar
\omega _c$  we can use the quantum limit approximation (QLA) \cite{zbt} which
 takes into account only diagonal pairing and ignores off-diagonal pairing
 completely. This is justified only if additional condition is fulfilled i.e.
 if the number of occupied Landau levels $n_{c}$ is less than $\approx
E_{F}/T_{c0}$,
where $E_{F}$ is the Fermi energy and $T_{c0}$ is the zero field transition
temperature. This is a situation that can readily be achieved in
low-carrier-density
systems by application of fields in the $1-30$ Tesla range. In lower fields,
where $n_{c}$ is large, off-diagonal terms have to be included as well:
 their number grows as $n_{c}^{2}$, while the number of diagonal terms grows as
$n_{c}$ and for sufficiently large $n_{c}\geq E_{F}/T_{c0}$ these terms will
eventually come to dominate.
For the moment, we ignore Zeeman splitting but we will show below how our
results can be generalized
to the case when the Zeeman effect is included.

Taking the order parameter in the form $\Delta(\vec{r})=\sum_{j}
\Delta_{j}(\vec{r})$
and after performing the BdG transformations (\ref{btransf}) we get
the following set of equations:
\begin{eqnarray}
E_
{k_{z},\vec{q}}^{N}
u_
{k_{z},\vec{q},n}^{N}
=(\frac{\hbar^{2}k_{z}^{2}}{2m}
+\hbar \omega _{c}(n+1/2)-\mu)u_
{k_{z},\vec{q},n}^{N}
+\sum_{m} \Delta_{nm}(\vec{q})v_
{k_{z},\vec{q},m}^{N}
\nonumber \\
-E_
{k_{z},\vec{q}}^{N}
v_
{k_{z},\vec{q},n}^{N}
=(\frac{\hbar^{2}k_{z}^{2}}{2m}
+\hbar \omega _{c}(n+1/2)-\mu)v_
{k_{z},\vec{q},n}^{N}
-\sum_{m} \Delta_{mn}(\vec{q})^{\ast} u_
{k_{z},\vec{q},m}^{N}
\label{bdg}
\end{eqnarray}
where $\Delta_{nm}(\vec{q})=\sum_{j} \Delta_{n,m}^{j}(\vec{q})$ is the matrix
element of $\Delta(\vec{r})=\sum_{j} \Delta_{j}(\vec{r})$ between electronic
states $(k_{z},\vec{q},n)$ and $(-k_{z},-\vec{q},m)$ and is given by:
\begin{eqnarray}
\Delta_{nm}^{j}(\vec{q})=\frac{\Delta_{j}}{\sqrt{2}}
\frac{(-1)^{m}}{2^{n+m}\sqrt{n!m!}}
\frac{(n+m)!}{(n+m-j)!}
\nonumber \\
\sum_{k}\exp{(i \pi \frac{b_{x}}{a}k^{2}+2ikq_{y}b_{y}
-(q_{x}+\pi k/a)^{2}l_{H}^{2})}
H_{n+m-j}[\sqrt{2}(q_{x}+\frac{\pi k}{a})l_{H}]
\label{diag}
\end{eqnarray}
$\Delta_{nm}^{j}(\vec{q})$ constitutes a magnetic lattice in
$\vec{q}$-space (i.e. a lattice which is invariant under MTG
transformations in $\vec{q}$-space) and belongs to the $(n+m-j)th$ Landau level
of charge $e/2$ in this space.
In the same way as in real space, the operator (\ref{pi}) can be constructed
in $\vec{q}$-space:
\begin{equation}
\Pi ^{\dagger}(\vec{q})=\frac{1}{2l_H}(-i\frac{\partial}{\partial q_{y}}
-\frac{\partial}{\partial q_{x}}+2q_{x}l_{H}^2)
\label{piq}
\end{equation}
 in order to obtain all $\Delta_{nm}^{j}(\vec{q})$ from
$\Delta_{00}^{j=0}(\vec{q})$
belonging to the lowest level in this space.
Behavior of $\Delta_{nm}^{j}(\vec{q})$ is very important for the excitation
spectrum and should be investigated in detail. The set of
$\Delta_{nm}^{j}(\vec{q})$
can be classified by the position and order of its zeros
: All $\Delta_{nm}^{j}(\vec{q})$ with $(n+m-j+4k)$ for quadratic and
$(n+m-j+6k)$ for
triangular lattices ($k$ is an integer) have the same set of zeros with the
similar dispersion around each zero. They differ considerably only far from
these points. Therefore, the contributions $\Delta_{j=4k}(\vec{r})$
($\Delta_{j=6k}(\vec{r})$)
with $k\neq 0$ for quadratic (triangular) lattice to $\Delta
(\vec{r})=\sum_{j}\Delta _{j}(\vec{r})$
have similar matrix elements around singular points to the matrix elements
$\Delta _{nm}^{j=0}(\vec{q})$ obtained for the order parameter entirely in the
lowest Landau level. Their inclusion does not bring any new qualitative feature
in the form of the excitation
spectrum and we
proceed by taking only $j=0$ contributions to $\Delta(\vec{r}) .

In very high magnetic fields ($\Delta \ll \hbar \omega _{c}$) we follow the QLA
and ignore the off-diagonal
pairing. Then, BdG equations (\ref{bdg}) can be solved analytically yielding
the quasiparticle excitation spectrum of the form:
\begin{equation}
E_
{k_{z},\vec{q},n}
=\pm\sqrt{[\frac{\hbar^{2}k_{z}^{2}}{2m}+\hbar \omega_{c}(n+1/2)-\mu]^{2}+
|\Delta_{nn}(\vec{q})|^{2}}
\label{spectrum}
\end{equation}
There are $n_{c}$ gapless branches in the above spectrum.
The gap $\Delta _{nn}(\vec{q})$ turns to zero on the Fermi surface at the set
of points $\{q_{j}=q_{yj}+iq_{xj}\}$ in the MBZ which are in direct
correspondence with the position of the vortices $\{z_{i}=x_{i}+iy_{i}\}$ e.g.
$q_{j}l_{H}=z_{i}/l_{H}$.
 While these zeros are of the first order, we have found that $n=1+2k$ branches
 for quadratic and $n=2+3k$ branches for triangular lattice have in addition
zeros of the second and third order, respectively.
Configuration of the zeros in the MBZ and corresponding vorticities are such as
to
preserve exactly one positive vorticity per unit cell of the order parameter
in real space. Figure 1. shows $E_{k_{F},\vec{q},0}$ branch in the spectrum
(\ref{spectrum}) of the triangular vortex lattice.

Lowering the magnetic field (but still in the region of $\Delta \le \hbar
\omega_{c}$ ), the number of occupied Landau levels grows and
the off-diagonal coupling becomes important (see the discussion above).
The Cooper pair are formed from the electrons in states $(k_{z},\vec{q},n)$
and $(-k_{z},-\vec{q},n\pm m)$ where $m\ll \Omega_{D}/\hbar \omega _{c}$
($\Omega_{D}$ is a Debye frequency).
Inclusion of the off-diagonal matrix elements $\Delta_{nm}(\vec{q})$ makes
solving the BdG equations (\ref{bdg}) a cumbersome problem
of diagonalizing the $2(n_{c}+M)\times 2(n_{c}+M)$ matrix that can be done
numerically.
Initially, it seems that the inclusion of the off-diagonal matrix elements
$\Delta_{nm}(\vec{q})$
destroys the gapless behavior of the excitation spectrum (\ref{spectrum}):
Off-diagonal matrix elements $\Delta_{nn+k}(\vec{q})$, where $k$ is an odd
integer, have zeros on the lattice dual to the lattice formed by zeros of
diagonal $\Delta_{nn}(\vec{q})$ and
there are no points in the MBZ where zeros of all matrix elements coincide.
However, our numerical results show that there are always
$n_{c}$ gapless branches in the excitation spectrum with the
position of zeros in the $(q_{x},q_{y})$ plane exactly the same as those found
within QLA. The role of off-diagonal matrix elements is to shift the value of
Fermi momentum $k_{Fn}$ at which the gapless behavior occurs in QLA to the new
value $k_{Fn}^{'}$
estimated from the condition:
\begin{equation}
\frac{\hbar ^{2}k_{Fn}^{'2}}{2m}-\frac{\hbar ^{2}k_{Fn}^{2}}{2m}
\approx \left( |\Delta
_{nn-k}(\vec{q_{j}})|^{2}-|\Delta_{nn+k}(\vec{q_{j}})|^{2}\right)/\hbar
\omega_{c} ,
\label{shift}
\end{equation}
 where $n$ is
the label of the Landau level. The shift (\ref{shift}) corresponds
to
the change in self-energies of the normal electronic Green functions by the
inclusion of the off-diagonal pairing. Figure 2. shows one of the gapless
branches in the excitation spectrum that
occurs at the Fermi momentum $k_{F3}^{'}$ determined from condition
(\ref{shift}) as $\hbar ^{2}k_{F3}^{'2}/2m\approx 0.85(\Delta ^{2}/\hbar \omega
_{c})$
for the case when all the Landau levels separated by $\hbar \omega _{c}$ and
$2\hbar \omega _{c}$ are coupled.
The gapless feature of the excitation spectrum is a direct consequence of the
behavior of the order parameter in
real space: Accommodating the magnetic field in the interior
of the superconductor, the order parameter assumes a non-uniform periodic
configuration
with exactly one flux quantum per unit cell. This topology  is reflected in the
 energy spectrum and results in the gapless behavior of the quasiparticle
excitation spectrum.
This result will hold as long as the region of interest in phase diagram is far
enough
from $H_{c1}(T)$ line so that the magnetic field inside the superconductor is
approximately uniform and as long as the temperature is low enough for the
quasiparticles to propagate coherently over many unit cells of the vortex
lattice.

If the magnetic field is lowered even further so that $\Delta \geq \hbar
\omega_{c}$,
mixing of Landau levels becomes very strong and the Landau level structure is
not very well defined anymore.
Monitoring the shift in $k_{Fn}$ as a function of $\Delta /\hbar \omega_{c}$,
we find that when $\Delta/\hbar \omega _{c}$ reaches the critical value
estimated from the relation:
\begin{equation}
\left( \frac {\Delta}{\hbar \omega_{c}}\right)_{critical} \approx \frac
{1}{2f_{nn+1}}
\label{critic}
\end{equation}
$n$-th and $n+1$-th gapless branches cross each other.
Increasing $\Delta /\hbar \omega _{c}$ (corresponds to decrease in magnetic
field) above the critical value (\ref{critic}) opens the gap at the Fermi
surface in this branch. The 'form factor' $f_{nn+1}$ ($n=0,2,4,...$ for the
quadratic and
$n=0,3,6,...$ for the triangular lattice) behaves as $f_{nn+1}\sim
\frac{1}{2^{n}n}$ for $n$ large,
so that the critical value of $\Delta /\hbar \omega_{c}$
increases as more and more Landau levels cross the Fermi surface.
The above branch crossing is similar to the behavior described in \cite{axel},
where the influence of an ordinary external periodic potential on a
2D electron system in a magnetic field was investigated.

In the discussion so far, we have ignored the Zeeman term $-g\mu
_{B}\hat{\sigma} \cdot {\bf H}(\vec{r})$
which is justified for lot of materials that have very small effective
g-factors. Now,
we show how our results can be generalize for the case when $g\neq 0$. The
interesting situation happens when $g\approx 2$ e.g. the Zeeman splitting is
closed to the cyclotron splitting  making the $n$th spin-up Landau level
nearly degenerate with the $(n+1)$-th spin-down
one. In this case the lattice of zeros found in $g=0$ case
makes a transition to the dual lattice described above for the off-diagonal
matrix
elements $\Delta_{nm}(\vec{q})$ with $n+m$ odd. Then, when $g\approx 4$ the
dual lattice transforms back to the original lattice, and so on.
 When $0 <  g < 2$  one should pair electrons with the momenta along the field
axis $k_{z}$ and $-k_{z}+q_{zn}$ for all possible $q_{zn}$ required to off-set
the
Zeeman splitting. In this case, one should consider a non-uniform (along the
field direction) order parameter of the form
$\Delta(x,y)exp(\sum_{n}q_{zn}\zeta)$,
where $\Delta(x,y)$ is given by (\ref{del}). The diagonalization of the BdG
equations (\ref{bdg}) is a very complicated task now due to the determination
of
the vectors $\vec{q_{zn}}$. Nevertheless, we still find the gapless points in
the excitation spectrum for the case where only few Landau levels are
occupied.
\\\\
{\bf III. THERMODYNAMIC PROPERTIES AND DENSITY OF STATES}

The strong dispersion around
gapless points at the Fermi surface leads to unusual temperature behavior of
the thermodynamic
functions at low temperatures.
Our calculations show that the heat capacity at low temperatures behaves as
:
\begin{equation}
c_{V}\approx \frac {k_{B}}{(2\pi l_{H})^{2}}(\frac {\Delta}{\hbar v_{F}})\frac
{N_{g}}{a_{n}^{2}}[(\frac {k_{B}T}{\Delta})^{3}+a_{n}^{\alpha -1}(\frac
{k_{B}T}{\Delta})^{\alpha }]
\label{heat}
\end{equation}
where $\alpha =2$ for the quadratic and $\alpha =5/3$ for  the triangular
lattice. $N_{g}$ is the
number of the gapless branches at some value of $\Delta/\hbar \omega_{c}$ (see
the discussion above). The $(k_{B}T/\Delta )^{\alpha }$ behavior in
(\ref{heat})
 is due to the presence of second (third) order zeros for
the quadratic (triangular) lattice in some of the branches of the energy
spectra,
while the $(k_{B}T/\Delta )^{3}$ behavior is a consequence of linear
dispersion around first order zeros in all off the gapless branches.
Coefficient
$a_{n}$ in (\ref{heat}) that measures slope around zero, depends strongly on
the chosen lattice symmetry but also
on the strength of the magnetic field.
 In lower fields more Landau levels
cross the Fermi surface and mix together making the slope around zero steeper
(e.g. increasing $a_{n}$, compare Fig. 1. and Fig. 2.). Furthermore, it was
mentioned before that in lower
fields the contribution to the order parameter from higher Landau levels
should be included as well. This will introduce wiggles in the region between
zeros in $\vec{r}$-space, making the slope around zero in the energy spectra
even steeper. Also, as the value of $(\frac {\Delta}{\hbar \omega_{c}})$
increases the number of gapless branches $N_{g}$ decreases, making the
algebraic temperature dependence in (\ref{heat}) weaker.

It is interesting to see how the superconducting density of states changes
from the standard BCS form due to the presence of zeros at the Fermi surface
in the quasiparticle excitation spectra. Our calculations show that the
low-energy
density of states per gapless branch of the spectrum has a behavior :
\begin{equation}
N_{s}(E)\approx N_{f}(0)[\frac {1}{a_{n}}(\frac {E}{\Delta})^{2}+
\frac {1}{a_{n}^{\alpha }}(\frac {E}{\Delta})^{\alpha }]
\label{dens}
\end{equation}
where $\alpha =1$ for the quadratic and $\alpha =2/3$ for the triangular
lattice.
$N_{f}(0)$ is the density of states of the free 3D system of electrons in a
magnetic field at the Fermi level. The term $(\frac {E}{\Delta})^{2}$ in
(\ref{dens})
comes from first order zeros while $(\frac {E}{\Delta})^{\alpha }$ is due to
second (third) order zeros in the gapless branches of the excitation spectrum
of the quadratic (triangular) vortex lattice.
\\\\
{\bf IV. THE TUNNELING PROPERTIES}

In this section we present
a theoretical study of various tunneling characteristics of a superconductor
in a high magnetic field exhibiting the gapless behavior described above.
We present the results for the triangular vortex lattice that is known to be
the lowest energy state in the mixed phase. First, we consider a simple problem
of
tunneling between two superconductors separated by a
thin insulating layer. The geometry is such that the tunneling occurs
primarily along the vortex lines. The tunneling problem was first studied by
Bardeen \cite{bardeen} and Cohen et.al. \cite{cohen} in the approximation of
the semiconductor band model where the chemical potentials of two
superconductors differ by the applied
voltage, i.e. $eV=\mu _{L}-\mu _{R}$. The potential drop $eV$
occurs in the insulating region between superconductors,
which is typically a metal oxide. Cohen et.al.
introduced the concept of a tunneling Hamiltonian:
\begin{equation}
H_{T}=\sum_{[k],[p]} T_{kp} c_{k}^{\dagger} c_{p} + h.c.~~~,
\label{tunnel}
\end{equation}
where $c_{k}$($c_{p}$) are set of operators describing electrons in a left
(right) superconductor.
 The tunneling matrix element $T_{kp}$
transfers particles through the insulating layer and is not spin dependent.
The tunneling in the superconductors will take place over a very narrow span
of energies around Fermi level, therefore it is adequate to treat the transfer
rate $T_{kp}$ as
 constant $T_{o}$ evaluated at $[k_{F}]$ and $[p_{F}]$. This is a proper
formalism  for the voltages in our problem ($eV\ll \Delta \approx 1$meV).

Following Ref.15, tunneling current through the
insulating layer between the superconductors is given by:
\begin {equation}
I(V,T)=2e\sum_{[k],[p]}|T_{kp}|^{2}\int_{-\infty }^{+\infty }\frac{d\varepsilon
}{2\pi } A_{R}(k,\varepsilon )A_{L}(p,\varepsilon +eV)
[n_{F}(\varepsilon )-n_{F}(\varepsilon +eV)]
\label{current}
\end{equation}

where $A_{R}(k,\varepsilon )$ and $A_{L}(p,\varepsilon +eV)$ are the spectral
 functions of right and left superconductor respectively.
$[k]=(\vec{k},k_{z},n)$ ($[p]=(\vec{p},p_{z},m)$) is a
set of quantum numbers representing the quasiparticle states in left (right)
superconductor and $n_{F}(\varepsilon )$ is a Fermi distribution.
 We anticipate that the most significant
contribution to the tunneling current for small voltages ($eV\ll \Delta $)
will come from the states around the gapless points in the quasiparticle
excitation spectrum. In the vicinity of these special points ($\vec{q_{i}},k
_{F}$)
at the Fermi surface the energy spectrum (\ref{spectrum}) can be approximated
as:
\begin{equation}
E(\vec{q},k_{z})\approx \sqrt{(\frac{E_{F}}{k_{F}}k_{z})^2+\Delta ^{2}a_{\alpha
}^{2}(q_{x}^{2}l^2+q_{y}^{2}l^2)^{\alpha }}
\label{energy}
\end{equation}
where $\alpha $ is the order of the zero, while the coefficient $a_{\alpha }$
measures slope around zero. $E_{F}$ is the Fermi level and $k_{F}$ is the
Fermi momentum. We are interested in the tunneling along the vortex lines i.e.
along the direction of magnetic field. For such a tunneling, in-plane quasi-
momenta are conserved up to the reciprocal lattice vector in the first MBZ.
The
quasiparticle from left with quasi-momentum $\vec{q}\approx \vec{q_{i}}$ can
tunnel in four equivalent states in the right superconductor with the same
contribution to the tunneling current.
After some algebra expression (\ref{current}) reduces to :
\begin{eqnarray}
I_{||}(V,T)\approx \frac{16\pi e}{\hbar
}N_{gR}N_{gL}|T_{o}|^{2}\sum_{\vec{q}}\sum_{k_{z},q_{z}>0}
[n_{F}(E_{\vec{q}k_{z}})-n_{F}(E_{\vec{q}q_{z}})]\delta (E_{\vec{q}q_{z}}-
eV-E_{\vec{q}k_{z}})
\nonumber\\
+[n_{F}(E_{\vec{q}q_{z}})-n_{F}(E_{\vec{q}k_{z}})]\delta(E_{\vec{q}k_{z}}-
eV-E_{\vec{q}q_{z}})
\label{parallel}
\end{eqnarray}
where we have assumed that there are $N_{gR}$ and $N_{gL}$ gapless branches in
 the excitation spectra of right and left superconductor respectively.

Differential conductance $\sigma=\partial I_{||}/\partial
V$ can be evaluated numerically from
(\ref{parallel}) for finite temperatures. For $T=0$ we can find the analytical
expression for
 differential conductance as :
\begin{equation}
\frac{\sigma (V,T=0)}{\sigma _{o}} = N_{gL}N_{gR}\left[ \frac{\Delta
_{R}a_{1R}}
{\Delta _{L}a_{1L}}(\frac{eV}{\Delta _{R}a_{1R}})^{2} B_{1}(\frac{\Delta
_{L}a_{1L}}
{\Delta _{R}a_{1R}})+(\frac{\Delta _{R}a_{3R}}{\Delta _{L}a_{3L}})
(\frac{eV}{\Delta _{R}a_{3R}})^{2/3}B_{3}(\frac{\Delta _{L}a_{3L}}{\Delta
_{R}a_{3R}})\right]
\label{sigma}
\end{equation}
where $\sigma _{o}$ is the tunneling differential conductance between two
 normal metals in a magnetic field and is given by :
\begin{equation}
\sigma _{o}=\frac{4\pi e^{2}}{\hbar } \frac{N_{1L}(0)}{2\pi l^{2}}
\frac{N_{1R}(0)}{2\pi l^{2}}|T_{o}|^{2}.
\label{sigmao}
\end{equation}
$N_{1L}(0)$ and $N_{1R}(0)$ are one-dimensional densities of states of a normal
 metal. $B_{\alpha }$ in (\ref{sigma}) are numerical factors associated with
the geometry of the vortex
lattice and the order of zeros $\alpha $ in the excitation spectrum
(\ref{energy}).
 They are equal to :
\begin{equation}
B_{\alpha }(Z)=2\pi ^{2}(2+\alpha )2^{\frac{2}{\alpha }}\frac{\Gamma ^{2}(
\frac{1}{\alpha })}{\Gamma (\frac{2}{\alpha})}\int_{0}^{\frac{1}{Z+1}}x^
{\frac{2}{\alpha }}F[\frac{1}{\alpha },\frac{1}{2};\frac{1}{\alpha
}+\frac{1}{2};
Z^{2}(\frac{x}{x-1})^2 ]dx
\label{geo}
\end{equation}
where $F(a,b;c;x)$ is the Hypergeometric function and $\Gamma (x)$ is the Gamma
 function.
Comparing (\ref{sigma}) and (\ref{dens}) we see that the differential
conductance has the same algebraic dependence on voltage as the density of
states on energy.
This dependence is a consequence of the presence of gapless points in the
quasiparticle excitation spectrum. $(eV)^{2}$ dependence in (\ref{sigma}) comes
from the presence of first order zeros while term $(eV)^{2/3}$ comes from third
order zeros in a spectrum of a triangular vortex lattice. At
finite temperatures this algebraic dependence will acquire the exponential tail
due to the thermal excitations of the quasiparticles over the Fermi surface.

In order to discuss the possible results of scanning-tunneling microscope
(STM) experiment in the presence of a vortex lattice, we need a model of the
 tunneling current between the microscope tip and surface of the superconductor
in a magnetic field. We will consider the same geometry of the experiment as
described for $S-N-S$ tunneling above, i.e. electrons from the tip can
tunnel only along vortex lines. The tunneling current will be related to the
spectral functions $A_{S}(\vec{r},\varepsilon )$ and $A_{N}(\vec{r},\varepsilon
)$
in a way described by the expression (\ref{current}) with the difference that
these functions now depend on the position of the tip $\vec{r}$. In our
 calculation we will assume that the presence of the surface does not affect
the quasiparticle wavefunctions of the superconductor. The differential
conductance is then given by :
\begin{equation}
\frac{\sigma (\vec{r},V,T)}{\sigma _{o}}=-\frac{2\pi l^{2}}{N_{1}(0)}
\sum _{q_{z}>0}\sum _{\vec{q}}\sum _{N} \left[
|u_{\vec{q}q_{z}}^{N}(\vec{r})|^{2}
 n_{F}^{'}(E_{\vec{q}q_{z}}^{N}-eV)+|v_{\vec{q}q_{z}}^{N}(\vec{r})|^{2}
n_{F}^{'}
(E_{\vec{q}q_{z}}^{N}+eV)\right]
\label{stm}
\end{equation}
where $n_{F}^{'}(E)$ is a derivative of a Fermi distribution and
$\vec{r}=(x,y)$
 is a position of a tip.
$u_{\vec{q}q_{z}}^{N}(\vec{r})=u_{\vec{q}q_{z}}^{N}\phi
 _{\vec{q}q_{z}}(\vec{r})$ and $v_{\vec{q}q_{z}}^{N}(\vec{r})=v_{\vec{q}q_{z}}^
{N}\phi _{\vec{q}q_{z}}(\vec{r})$ are the solutions of BdG
equations (\ref{bdg}) for superconductor in a magnetic field.
Sum $\sum _{N}$ in (\ref{stm})
goes over the gapless branches of the quasiparticle excitation spectrum, since
the most important contribution to the tunneling current at small bias voltages
comes from the gapless
points at the Fermi surface.

Figure 3. shows the differential conductance (\ref{stm}) at zero temperature as
a function of the
bias voltage when the microscope tip is at the position of the vortex.
 For small tunneling voltages it has the algebraic voltage
dependence, reflecting the behavior of the low-energy density of states
(\ref{dens}).
It has a peak for the value of the voltage approximately equal to
the maximum value of the order parameter.
 Note that this result differs from the one obtained in Ref. 5. for the
isolated
 vortex case where the conductance is strongly enhanced at zero bias.
 It is important to understand that our calculation is done for the
fundamentally different physical situation that the one considered in Ref. 5.
In high magnetic fields vortices are very close to each other so that at low
temperatures quasiparticles can propagate coherently over many unit cells.
This coherent propagation will influence all the properties of the
superconductor as we have shown in this paper.
 In lower magnetic fields, where the electronic mean free path is much shorter
than the separation of vortices, the quasiparticle states are effectively
bound to a single vortex. The relationship between these two limits is
essentially that of a single impurity problem versus an ordered lattice of
scatterers.
 Figure 4. shows how the differential
 conductance (\ref{stm}) depends on a tip position for the fixed value of
a biased voltage $V/\Delta = 0.2$. This figure has a six-fold symmetry pattern
of a triangular vortex lattice. Differential conductance has maxima at the
positions of the vortices.
 The conductance decreases as the tip
moves away from the vortex and reaches its minimum value at the positions where
 the order parameter has its maxima (these points form the hexagonal lattice).
\\\\
{\bf V. CONCLUSION}

In this paper we have shown that the BCS theory of a superconductor in a
magnetic field can be solved exactly leading to the gapless behavior of the
quasiparticle excitation spectrum in high magnetic fields.
We have also discussed the mechanism of gap opening in the lower fields. As a
result of the strong dispersion around gapless points, we have found the
algebraic behavior of the heat capacity at low temperatures as well as density
of states at low energies.
The differential conductance of an $S-N-S$ junction and differential
conductance
 between STM tip and the superconductor are found to have an algebraic
dependence on biased voltage. This unusual behavior of type-II superconductor
in
 high magnetic field can be in principle detected in suitably designed
experiments at low temperatures.
\\\\
{\bf ACKNOWLEDGMENTS}

Authors would like to thank A. V. Andreev for numerous useful discussions.

\newpage
\begin{figure}
\caption{$E_{k_{F},\vec{q},0}$ branch in the spectrum (11) of  the triangular
vortex lattice when $\Delta /\hbar \omega _{c}=1/20$.}
\label{fig 1}
\end{figure}

\begin{figure}
\caption{$E(k_{F3}^{'},\vec {q})$ branch in the quasiparticle excitation
spectrum of the triangular vortex lattice when $\Delta /\hbar \omega _{c}=1/5$
(off-diagonal pairing is included).}
\label{fig 2}
\end{figure}

\begin{figure}
\caption{The differential conductance as a function of a bias voltage when
the STM tip is at the position of the vortex (in arbitrary units).}
\label{fig 3}
\end{figure}

\begin{figure}
\caption{The differential conductance as a function of the position of a
STM tip for $V/\Delta = 0.2$. Full circles represent minima of differential
conductance while diamonds show the maxima. }
\label{fig 4}
\end{figure}

\end{document}